 \documentclass[twocolumn,showpacs,amsmath,prl,amssymb,showkeys,superscriptaddress,aps,nolongbibliography,nofootinbib]{revtex4-1}

\pdfoutput=1

\usepackage{graphicx}
\usepackage{dcolumn}
\usepackage{mathtools}
\usepackage{scalerel}
\usepackage{bm}
\usepackage{upgreek}
\usepackage{mathrsfs}
\usepackage{float}

  \usepackage{array}
 \usepackage{verbatim}
  \usepackage{amstext}
  \usepackage{amsmath}
  \usepackage{amssymb}
  \usepackage{slashed}
  \usepackage[bbgreekl]{mathbbol}
  \usepackage{indentfirst}
  \usepackage{xspace}
 \usepackage[T1]{fontenc}
 \usepackage[latin9]{inputenc}
 \usepackage{color}
 \PassOptionsToPackage{normalem}{ulem}
 \usepackage{ulem}
 \usepackage[unicode=true,pdfusetitle,
  bookmarks=true,bookmarksnumbered=false,bookmarksopen=false,
  breaklinks=false,pdfborder={0 0 1},backref=false,colorlinks=true]
  {hyperref}
 \hypersetup{
  citecolor=blue}

\newcommand{\Pmueb}     	 {P_{\bar{\mu} \bar{e}}}
\newcommand{\Pmue}     	 {P_{\mu e}}

\newcommand{\delcp}         {\delta_{cp}}

\newcommand{\tc}	{\theta_{23}}
\newcommand{\ldm}	{\Delta m_{31}^2}
\newcommand{\sdm}	{\Delta m_{21}^2}
\newcommand{\axem}	{a^X_{e \mu}}
\newcommand{\axet}	{a^X_{e \tau}}
\newcommand{\axmt}	{a^X_{\mu \tau}}
\newcommand{\cxyem}	{c^{XY}_{e \mu}}
\newcommand{\cxyet}	{c^{XY}_{e \tau}}
\newcommand{\cxymt}	{c^{XY}_{\mu \tau}}

\newcommand{\axab}	{a^X_{\alpha \beta}}
\newcommand{\cxyab}   {c^{XY}_{\alpha \beta}}

\newcommand{\phxab}	{\phi^X_{\alpha \beta}}
\newcommand{\phxyab}   {\phi^{XY}_{\alpha \beta}}

\makeatother

\begin{document}

\title{Octant Ambiguity in the Presence of Non-isotropic Lorentz Invariance Violation}

\newcommand{\cusb}{Department of Physics,
Central University of South Bihar, Gaya 824236, India}

\newcommand{\bhu}{Department of Physics, Institute of Science,
Banaras Hindu University,
Varanasi 221005, India.}

\newcommand{\corrsks}{saurabhshukla@cusb.ac.in}
\newcommand{\corrls}{lakhwinder@cusb.ac.in}
\newcommand{\corrskm}{shashank@cusb.ac.in}

\author{ Saurabh~Shukla} \altaffiliation{ \corrsks }\affiliation{ \cusb }\affiliation{ \bhu }
\author{ Shashank~Mishra} \altaffiliation{ \corrskm }\affiliation{ \cusb }\affiliation{ \bhu }
\author{ Lakhwinder~Singh }  \altaffiliation{\corrls}\affiliation{ \cusb }
\author{ Venktesh~Singh }   \affiliation{ \cusb }

\date{\today}

\begin{abstract}
Global analyses of neutrino data suggest that the mixing angle $\theta_{23}$ is likely to be nonmaximal with two closely matched solutions emerging: one representing a smaller angle ($\theta_{23}$ < $\pi/4$) and the other a larger angle ($\theta_{23}$ > $\pi/4$). This ambiguity, known as the octant ambiguity of $\tc$, presents a significant challenge in neutrino research and is a primary objective of future long-baseline experiments. 
In this study, for the first time, we explore how non-isotropic Lorentz violation affects measurements of mixing angle $\theta_{23}$, with a particular emphasis on sidereal effects in the Deep Underground Neutrino Experiment.
Our findings reveal that ability of DUNE to resolve the octant ambiguity is significantly compromised in the presence of the $\cxyet$ parameter. 
Furthermore, we demonstrate that LIV exacerbates the degeneracy between the Dirac CP-phase $\delcp$ and $\theta_{23}$.
 \end{abstract}

\maketitle
{\bf {\em Introduction:}} Neutrino oscillation, which involves neutrinos changing between three flavors, provides evidence that neutrinos have mass and challenges the Standard Model's assumption of massless neutrinos.
 Within the standard three-neutrino oscillation framework, key parameters include the three mixing angles $\theta_{12}$, $\theta_{13}$, and $\theta_{23}$, as well as the two mass squared differences, $\Delta m^{2}_{31} = m^{2}_{3} - m^{2}_{1}$ and $\Delta m^{2}_{21} = m^{2}_{2} - m^{2}_{1}$, along with the Dirac CP-phase $\delcp$. 
Current research focuses on precisely determining crucial parameters such as the octant of $\theta_{23}$, the leptonic CP-phase $\delcp$, and the sign of the atmospheric mass squared difference $\Delta m^{2}_{31}$ (Normal Hierarchy (NH): $\Delta m^{2}_{31} > 0$, Inverted Hierarchy (IH): $\Delta m^{2}_{31} < 0$)~\cite{Esteban:2020cvm}. Various studies indicate potential degeneracies among these parameters ~\cite{Barger:2001yr, Sugama:2023sjf, Ghosh:2015ena}. 
Understanding the octant of the neutrino mixing angle $\theta_{23}$ is crucial from a phenomenological perspective, as it affects important aspects such as the neutrino mass hierarchy and CP violation due to parameter degeneracy in their estimation \cite{Fogli:1996pv, Barger:2001yr, Ghosh:2015ena, Minakata:2001qm, Hiraide:2013ju2006vh, Burguet-Castell:2002ald, Agarwalla:2013ju, Machado:2013kya, Minakata:2013eoa, Chatterjee:2013qus}. As a result, current experimental efforts are focused on accurately measuring these parameters.

Global neutrino data suggests that the mixing angle $\theta_{23}$ is likely not maximal, leading to two nearly equal possibilities: the lower octant (LO), where $\theta_{23}$ less than $\pi/4$, and the higher octant (HO), where $\theta_{23}$ greater than $\pi/4$. Determining the correct octant $\theta_{23}$ is crucial due to its significant implications for theories of neutrino masses and mixing \cite {Mohapatra:2006gs, Albright:2006cw, Altarelli:2010gt, King:2014nza, King:2015aea}. Various theoretical frameworks underscore the importance of resolving the octant, including $\mu \leftrightarrow \tau$ symmetry ~\cite{Fukuyama:1997ky, Mohapatra:1998ka, Lam:2001fb, Harrison:2002et, Kitabayashi:2002jd, Grimus:2003kq, Koide:2003rx, Mohapatra:2005yu}, $A_4$ flavor symmetry \cite {Ma:2002ge, Ma:2001dn, Babu:2002dz, Grimus:2005mu, Ma:2005mw}, quark-lepton complementarity \cite {Raidal:2004iw, Minakata:2004xt, Ferrandis:2004vp, Antusch:2005ca}, and neutrino mixing anarchy \cite{Hall:1999sn, deGouvea:2012ac}.

Long-baseline (LBL) accelerator experiments are a powerful tool for identifying the $\theta_{23}$ octant. These experiments utilize both the $\nu_\mu \to \nu_e$ appearance channel and the $\nu_\mu \to \nu_\mu$ disappearance channel to enhance sensitivity to the octant of $\theta_{23}$. While the $\nu_\mu \to \nu_\mu$ survival probability, which depends primarily on $\sin^2 2\theta_{23}$, can reveal deviations from maximal mixing, it does not provide information about the octant. In contrast, the $\nu_\mu \to \nu_e$ appearance probability, which varies with $\sin^2 \theta_{23}$, is sensitive to the octant and can help differentiate between the LO and HO. By combining results from both channels, experiments can obtain a more comprehensive understanding of $\theta_{23}$.

Most studies on octant sensitivity have been conducted within the conventional framework of three-flavor neutrino oscillations \cite{Agarwalla:2013ju, Agarwalla:2013hma, Bass:2013vcg, Bora:2014zwa, Das:2014fja, Nath:2015kjg}. However, there are substantial reasons to consider that the Standard Model may be insufficient in explaining all aspects of neutrino behavior. Non-standard effects in neutrino oscillations could significantly impact the precision of $\theta_{23}$ measurements \cite{KumarAgarwalla:2019gdj, Agarwalla:2016xlg, Agarwalla:2016fkh, PhysRevD.108.095050}. One such effect is Lorentz invariance violation (LIV), which explores deviations from the fundamental symmetries predicted by the Standard Model. According to the Standard Model Extension, LIV could alter neutrino oscillation probabilities, potentially detectable through sidereal variations, daily modulations caused by Earth's rotation and its motion relative to distant celestial objects \cite{Diaz:2009qk, Kostelecky:2003xn}. In the context of long-baseline (LBL) experiments, where the neutrino beam is subject to such daily modulations, LIV effects may significantly influence the accuracy of standard neutrino oscillation parameter measurements and offer crucial insights into phenomena beyond the Standard Model \cite{Shukla:2024fnw}.

In this study, we examine the impact of non-diagonal non-isotropic Lorentz violation (LIV) parameters on the capabilities of the Deep Underground Neutrino Experiment (DUNE)~\cite{DUNE:2020lwj}. We highlight that the measurement of the $\theta_{23}$ octant could be jeopardized by the presence of these LIV parameters. Our analysis reveals a degeneracy between the $\theta_{23}$ octant and $\delta_{cp}$ phase that does not arise in the standard oscillation framework ~\cite{DUNE:2020jqi}. Additionally we also investigate how varying the strength of LIV parameters affects octant sensitivity and discuss the significant features observed in the sensitivity analysis for different LIV parameters. It is crucial to recognize that the precise determination of the $\theta_{23}$ octant in long-baseline experiments may encounter challenges due to the influence of new-physics scenarios.

{\bf {\em Formalism:}} Explicitly, one can write the Hamiltonian for the neutrino propagation in the matter as,
\begin{align}
  H_{\alpha\beta} &= \frac{1}{2E} \left[
           U_{\alpha j} {diag} [0, \sdm, \ldm]_{jk}
          (U^\dag)_{k \beta} + (\tilde{V}_{\rm MSW})_{\alpha\beta}
       \right].
  \label{eq:Hamiltonian}
\end{align}
Here, $\alpha$ and $\beta$ corresponds to the neutrino flavours. U is Pontecorvo-Maki-Nakagawa-Sakata (PMNS) matrix which is parameterized in terms of standard oscillation parameters and  $(\tilde{V}_{\rm MSW})$ is matter effect potential.

 The LIV-Hamiltonian ($H_{\rm LIV}$) for neutrino-neutirno 
 mixing is given by ~\cite{Kostelecky:2003xn}
\begin{equation}
  (\mathcal{H}_{LIV})_{\alpha\beta}  = \frac{1}{E}[(a_{L})^{\mu} p_{\mu}  - (c_{L})^{\mu\nu} p_{\mu} p_{\nu}]_{\alpha \beta}.
  \label{hLIV}
\end{equation}
The LIV Hamiltonian encompasses two categories of parameters: the CPT-violating LIV coefficients, denoted as $(a)_L^\mu$, and the CPT-conserving LIV coefficients, denoted as $(c)_L^{\mu\nu}$. Both $(a)_L^\mu$ and $(c)_L^{\mu\nu}$ are $3 \times 3$ complex matrices, representing LIV coefficients with mass dimensions 1 and 0, respectively.
In the case of antineutrino-antineutrino mixing, $(a_{L})^{\mu}$ becomes $- ((a_{L})^{\mu})^{*}$ and $(c_{L})^{\mu\nu}$ becomes $ ((c_{L})^{\mu\nu})^{*}$.
In the subsequent sections of this paper, these parameters will be referred to without the 'L' subscript.

The detection of non-isotropic LIV relies on observing changes in the neutrino beam direction, which can be tracked by monitoring Earth's rotation. This rotation is measured using distant stars as reference points, a method known as sidereal rotation. 
For terrestrial experiments, the standard inertial frame used is the Sun-centred celestial-equatorial frame (SCCEF), characterized by coordinates $(X, Y, Z, T)$ \cite{PhysRevD.66.056005}. In this frame, both the neutrino source and detector rotate with an angular frequency approximately equal to $2\pi$ divided by 23 hours and 56 minutes.
The sidereal time dependence of the effective Hamiltonian can be explicitly represented using directional factors~\cite{PhysRevD.109.075042, Kostelecky:2004hg}.
The orientation of the neutrino beam and the position of the detector are characterized by the directional factors \((N^X, N^Y, N^Z)\).
These factors are related to the Zenith angle $\theta$, which measures the angle between the beam and the vertical upward direction; the bearing angle $\phi$, which indicates the angle between the beam and the south direction measured towards the east; and the colatitude of the detector $\chi$~\cite{Kostelecky:2004hg}.
\begin{equation}
  \begin{split}
  N^{X} &= \cos\chi\sin\theta\cos\phi + \sin\chi\cos\theta,\\
  N^{Y} &= \sin\theta\sin\phi,\\
  N^{Z} &= -\sin\chi\sin\theta\cos\phi + \cos\chi\cos\theta, \\
  \end{split}
  \label{orientationbeam}
\end{equation}
The LIV coefficients $(a)_{\alpha \beta}^{\mu}$ are controlled solely by the baseline length, while the LIV coefficients, $(c)_{\alpha \beta}^{\mu\nu}$, are affected by both the baseline length and the neutrino energy.
In this study, we focus on six specific parameters: $a^{X}_{e\mu}$, $a^{X}_{e\tau}$, $a^{X}_{\mu\tau}$, $c^{XY}_{e\mu}$, $c^{XY}_{e\tau}$, and $c^{XY}_{\mu\tau}$. In the subsequent sections, the magnitude of the LIV parameters is denoted as $\axab$ and $\cxyab$, while the phase of these parameters is represented as $\phxab$ and $\phxyab$, respectively. For current limits on LIV parameters in the neutrino sector, refer to ~\cite{RevModPhys.83.11}.

In the presence of LIV parameters, the difference in oscillation probability can be expressed as the sum of the standard interaction contribution and the LIV contribution, up to leading-order terms, as mentioned in references ~\cite{PhysRevD.109.075042,Liao:2016hsa,Chaves:2018sih,Yasuda:2007jp}:
\vspace{1mm}
\begin{equation}
 \begin{split}
\Delta P_{\mu e} &\simeq \Delta P_{\mu e}(\text{SI}) + \Delta P_{\mu e}(e \mu) + \Delta P_{\mu e}(e \tau),\\
\Delta P_{\mu \mu} &\simeq \Delta P_{\mu \mu}(\text{SI}) + \Delta P_{\mu e}(\mu \tau).\\
 \end{split}
 \label{eq:DelPmeliv}
\end{equation}

It is noteworthy that in appearance channels, only $(e \mu)$ and $(e \tau)$ type parameters appear up to leading order, while $(\mu \tau)$ type parameters appear in the disappearance channel. 
To resolve the degeneracy, a measurement must differentiate between the oscillation probabilities associated with the two degenerate solutions. The goal of the experiment is to identify the octant of $\theta_{23}$ by detecting differences in probabilities between the true octant and the test octant as
\vspace{1mm}
\begin{equation}
\Delta P = P_{\mu e}^{\text{true}} (\theta_{23}^{\text{true}}, \delta_{cp}^{\text{true}}, \phi^{\text{true}})
           - P_{\mu e}^{\text{test}} (\theta_{23}^{\text{test}}, \delta_{cp}^{\text{test}}, \phi^{\text{test}}).
  \label{eq:oc}
\end{equation}

It is important to note that $\Delta P_{\mu e}$ is influenced not only by the standard CP phase $\delta_{cp}$ but also by the additional dynamical CP phase $\phi_{e\mu}$ or $\phi_{e\tau}$ related to Lorentz-violating (LIV) effects. This extra degree of freedom could potentially decrease the sensitivity to the octant of $\theta_{23}$.

This study analyzes the impact of individual parameters, focusing on one non-zero parameter at a time. 
It is crucial to emphasize that only parameters with strengths exceeding $10^{-24}$ will significantly affect the sensitivity (for probabilistic formalism see ~\cite{Agarwalla:2016fkh, KumarAgarwalla:2019gdj}). Each parameter is assigned a strength of $5.0 \times 10^{-23}$ for this analysis.

For numerical simulation of events for DUNE, the GLoBES package~\cite{Huber:2004ka, Huber:2007ji} is utilized with appropriate modifications specified in the latest configuration file provided by the collaboration. Simulations are carried out over 5 years in both neutrino and antineutrino modes. The experimental setup is based on the details outlined in the DUNE Technical Design Report (TDR) ~\cite{DUNE:2021cuw}, and the analysis includes both disappearance and appearance channels.
DUNE FD and beam orientation details are specified by  co-latitude ($\chi$)  of $48.3793^{\circ} $, zenith angle ($\theta$ ) of $84.26^{\circ} $, and  bearing angle ($\phi $ ) of $204.616^{\circ} $. The benchmark values for the standard oscillation parameters are: 
$\theta_{12} = 33.48^\circ$, $\theta_{13} = 8.5^\circ$, $\theta_{23} = 45.0^\circ$,
$\delta_{cp} = 195.0^\circ$, $\Delta m^{2}_{21} = 7.55 \times 10^{-5} \, \text{eV}^2$,
$\Delta m^{2}_{31} = 2.50 \times 10^{-3} \, \text{eV}^2$.
For the sensitivity calculation, we use Poisson-likelihood chi-square statistics, as detailed in Ref.~\cite{Baker:1983tu, Shukla:2024fnw}.
For the marginalization range,  $\theta_{23}$ varies from $41.0^\circ$ to $52.0^\circ$,  $\Delta m^{2}_{31}$ parameter is considered for both normal and inverted mass hierarchies, while $\Delta m^{2}_{21}$ is varied within its $1\sigma$ range. Additionally, all the phases $\delta_{cp}$, $\phi^X_{\alpha \beta}$, and $\phi^{XY}_{\alpha \beta}$ are varied from $[-\pi, \pi]$. 
    
\begin{figure}[!]
      \centering
      \includegraphics[height=0.20\textwidth,width=0.40\textwidth]{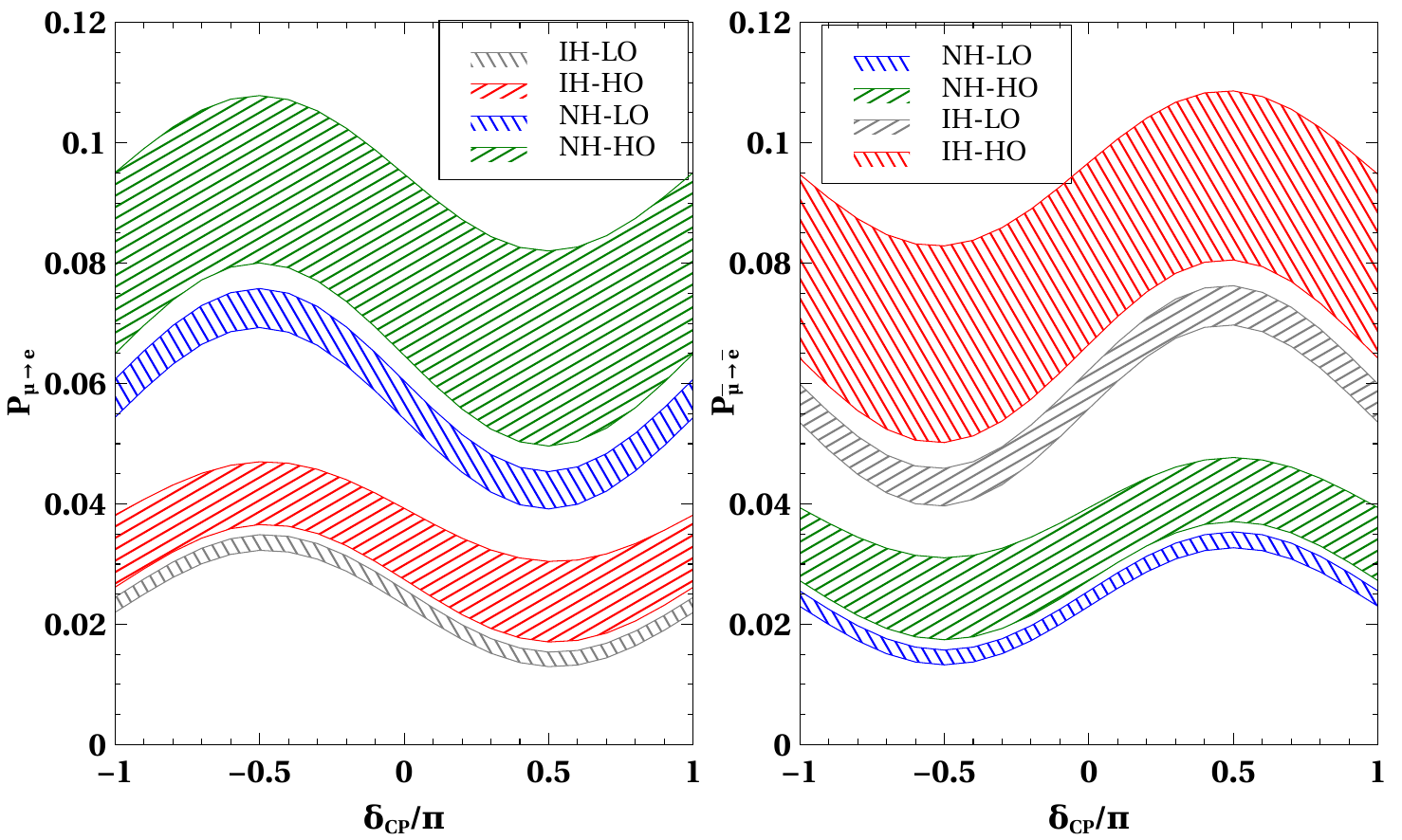}
    \caption{Oscillation probability $\Pmue$ ($\Pmueb$) as a function $\delta_{cp}$ at $E$ = 2.5 GeV in the SM case showing neutrino probabilities in the left panel and antineutrino probabilities in the right panel. }
    \label{probsm}
    \end{figure}

{\bf {\em Results:}} 
Figure ~\ref{probsm} presents the appearance oscillation probabilities for neutrinos (left) and antineutrinos (right) at an energy of 2.5 GeV as a function of the CP-violating phase $\delta_{cp}$, comparing various scenarios: Normal Hierarchy - Lower Octant (NH-LO), Normal Hierarchy - Higher Octant (NH-HO), Inverted Hierarchy - Lower Octant (IH-LO), and Inverted Hierarchy - Higher Octant (IH-HO). The figure ~\ref{probsm} reveals that neutrino appearance probabilities are notably higher in the Normal Hierarchy (NH) case due to stronger matter effects, whereas antineutrino probabilities are elevated in the Inverted Hierarchy (IH) case, attributed to weaker matter effects. Additionally, the peaks of the oscillation probabilities occur in the lower half of $\delta_{cp}$ for neutrinos and in the upper half for antineutrinos. The higher octant scenarios consistently show greater probabilities due to the contribution of the $\sin^2{\theta_{23}}$ term. The absence of overlapping probability bands across different scenarios indicates no degeneracy in the standard case.

\begin{widetext}
	    
    \begin{figure}[H]
      \centering
      \includegraphics[height=0.40\textwidth,width=0.90\textwidth]{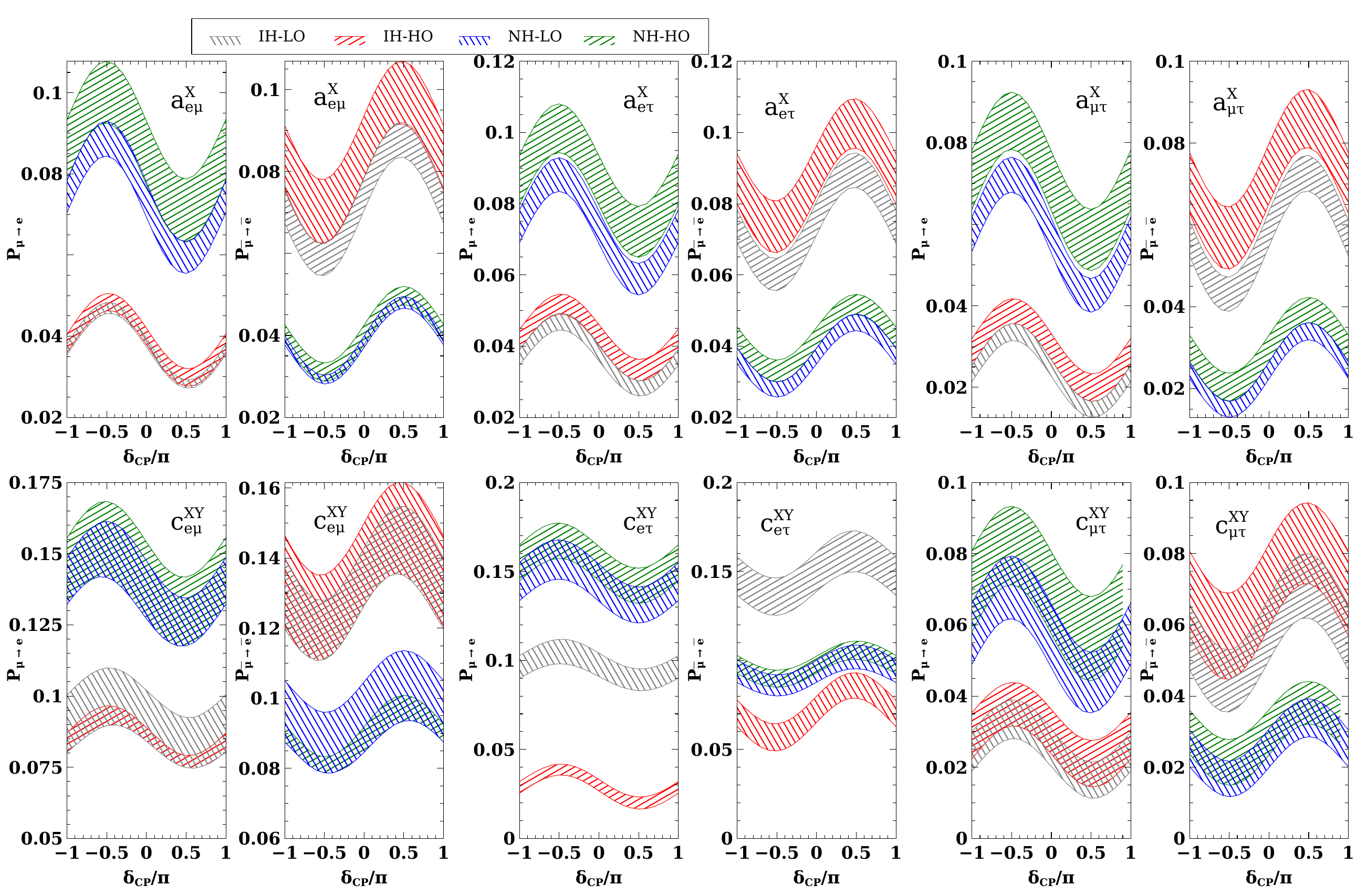}
    \caption{ Probability of $\mu \to e$ (or $\bar{\mu} \to \bar{e}$) appearance as a function of $\delta_{cp}$ at $E = 2.5$ GeV in the SM+LIV scenario. The top (Bottom) six panels show $a^{X}_{e\mu}$, $a^{X}_{e\tau}$ $a^{X}_{\mu\tau}$ ( $c^{XY}_{e\mu}$, $c^{XY}_{e\tau}$, $c^{XY}_{\mu\tau}$) for the neutrino and antineutrino case. Different mass hierarchies and octants are shown with color bands.}
    \label{probliv}
    \end{figure}
    
        \end{widetext}

Figure ~\ref{probliv} introduces LIV parameters set to $5.0 \times 10^{-23}$. With these LIV parameters, the probability curves exhibit significant overlap, suggesting potential degeneracies. Notably, overlaps are observed in the bands for $a^{X}_{e\mu}$ , $c^{XY}_{\mu \tau}$ and $c^{XY}_{e\tau}$ across different octants within both hierarchies. Despite this, different mass hierarchy scenarios within a given octant remain distinguishable. The introduction of LIV parameters thus adds complexity to the measurement of octant effects, which could impact the interpretation of results from experiments such as DUNE.

\begin{figure}[!]
      \centering
       \includegraphics[height= 0.25\textwidth,width=0.45\textwidth]{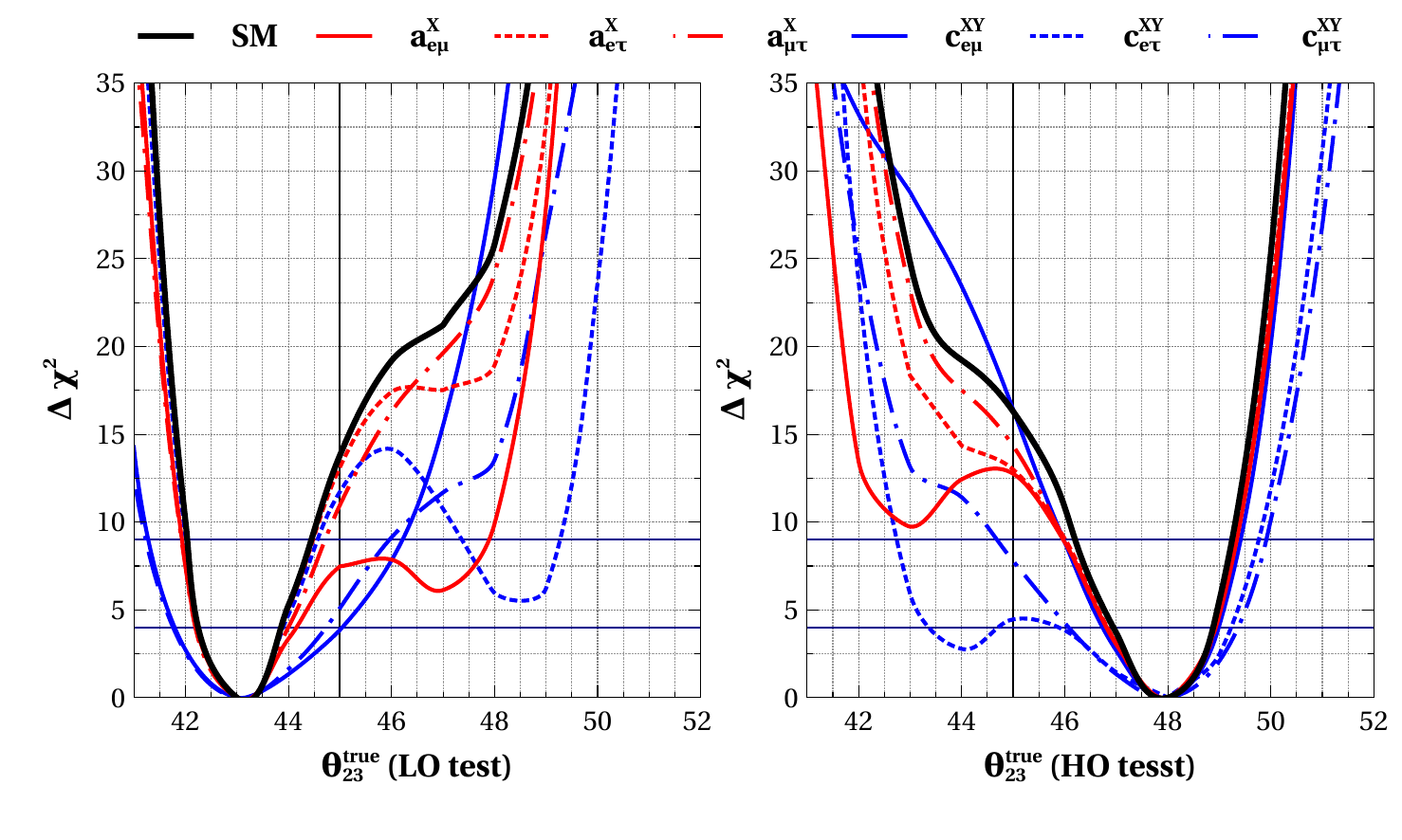}
      \caption{Octant sensitivity for the DUNE as a function of true $\theta_{23}$ values. The black line denotes the SM case, while different line styles indicate SM+LIV ($\axab$ / $\cxyab$) scenarios in pink and blue. The left (right) panel shows sensitivity with the lower octant (higher octant) as the true value.}  
         \label{th23sensitivity}
    \end{figure}

We now examine the sensitivity of determining the octant $\theta_{23}$ as a function of true value of $\theta_{23}$. The parameters $\Delta m^2_{31}$ and $\theta_{13}$ are marginalized over both their true and test values. Additionally, all phases are marginalized across the range $[-\pi, \pi]$ in both the true and test cases.
In Fig.~\ref{th23sensitivity}, we show the sensitivity of DUNE to the mixing angle $\theta_{23}$. The figure is divided into two panels: the left panel corresponds to the Low Octant (LO) with true $\theta_{23} = 43^\circ$, while the right panel corresponds to the High Octant (HO) with true $\theta_{23} = 48^\circ$. In the absence of LIV effects, DUNE can accurately determine the true value of $\theta_{23}$ with up to $3\sigma$ confidence, and there is potential to improve this sensitivity to $5\sigma$, surpassing current levels of uncertainty.
However, the inclusion of LIV parameters specifically $a^{X}_{e\mu}$, $c^{XY}_{\mu \tau}$, and $c^{XY}_{e\tau}$ leads to a significant decrease in DUNE's sensitivity to the octant of $\theta{23}$. 
This effect is further illustrated in Fig.~\ref{probliv}, where overlapping probability distributions for different octants become evident due to the presence of these LIV parameters.
The presence of these LIV parameters effectively undermines DUNE's ability to distinguish between the low and high octants of $\theta_{23}$, which could, in turn, affect the precision of measurements for other oscillation parameters such as $\delta_{cp}$.
 
\begin{widetext}
	    
    \begin{figure}[H]
      \centering
       \includegraphics[height= 0.35\textwidth,width=0.85\textwidth]{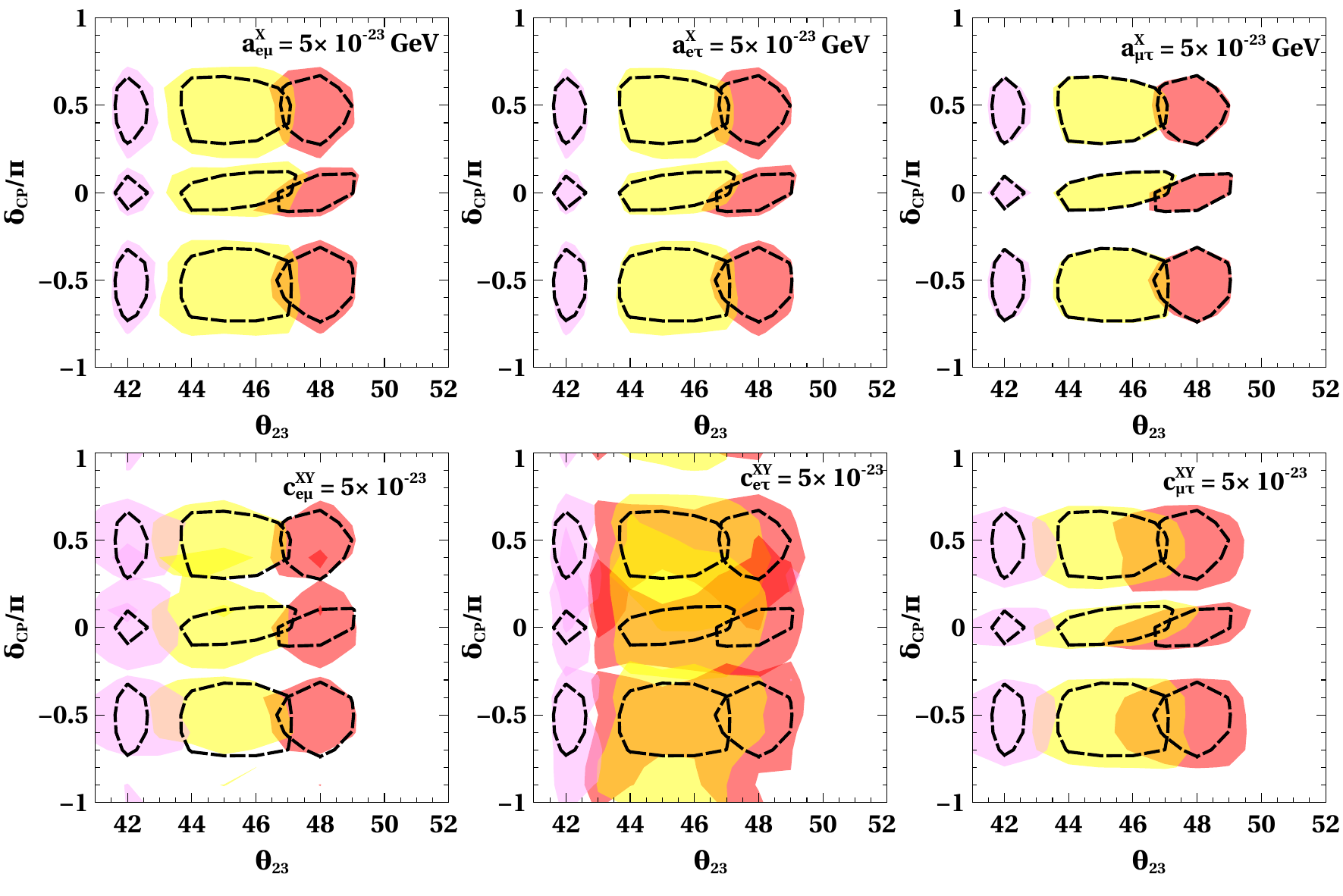}
     \caption{The two-dimensional $90\%$ confidence regions in the $\theta_{23}$ vs. $\delta_{cp}$ plane for DUNE after 10 years of equal neutrino and antineutrino exposure. The dotted line represents the SM, and SM+LIV scenarios are shown with pink, yellow, and red regions for true $\theta_{23}$ values of $\{42^\circ, 45^\circ, 48^\circ \}$, with $\delcp$ values of $\{\pi, 0, -\pi\}$. True values are indicated by a star.}
     \label{Corelationth23delcp}
    \end{figure}
    
        \end{widetext}

 Furthermore, we explore the correlation between $\delta_{cp}$ and $\theta_{23}$, considering specific test values for both parameters in the presence of LIV.
 Here, test parameters $\Delta m^2_{31}$ and $\theta_{13}$ are marginalized over their respective  ranges, along with the LIV phases $\phi^X_{\alpha \beta}$ and $\phi^{XY}_{\alpha \beta}$ over $[-\pi, \pi]$ in both the true and test cases. True $\theta_{23}$ and true $\delcp$ are varied in their respective ranges.   
In Figure \ref{Corelationth23delcp}, we depict the $90\%$ confidence level $\Delta \chi^2$ allowed region within the $\theta_{23}$ and $\delta_{cp}$ plane, derived from (5+5) years of equal running. Our analysis delves into the correlation between $\theta_{23}$ and $\delta_{cp}$ across different combinations of test values, namely $\theta_{23}$ as $\{42^\circ, 45^\circ, 48^\circ\}$ and $\delta_{cp}$ as $\{-\frac{\pi}{2}, 0, +\frac{\pi}{2}\}$. The dotted line delineates the SM scenario, whereas the shaded region denotes the influence of LIV. Notably, the resolution in $\theta_{23}$ deteriorates at its maximal value but improves as it deviates from this extremum in either octant across all scenarios. Conversely, the resolution in $\delta_{cp}$ remains robust. In the absence of LIV, the SM case exhibits no significant degeneracy between $\theta_{23}$ and $\delta_{cp}$ (ref). This condition persists in the presence of $\axem$, $\axet$, $\axmt$, and $\cxyem$. However, degeneracy emerges with $\cxyet$ and $\cxymt$, particularly affecting $\theta_{23}$. Notably, the lower octant value for both parameters remains non-degenerate, while the maximal and higher values exhibit overlap in their shaded regions, indicating degeneracy across all chosen true values of $\delta_{cp}$.

        \begin{figure}[!]
      \centering
       \includegraphics[height= 0.25\textwidth,width=0.40\textwidth]{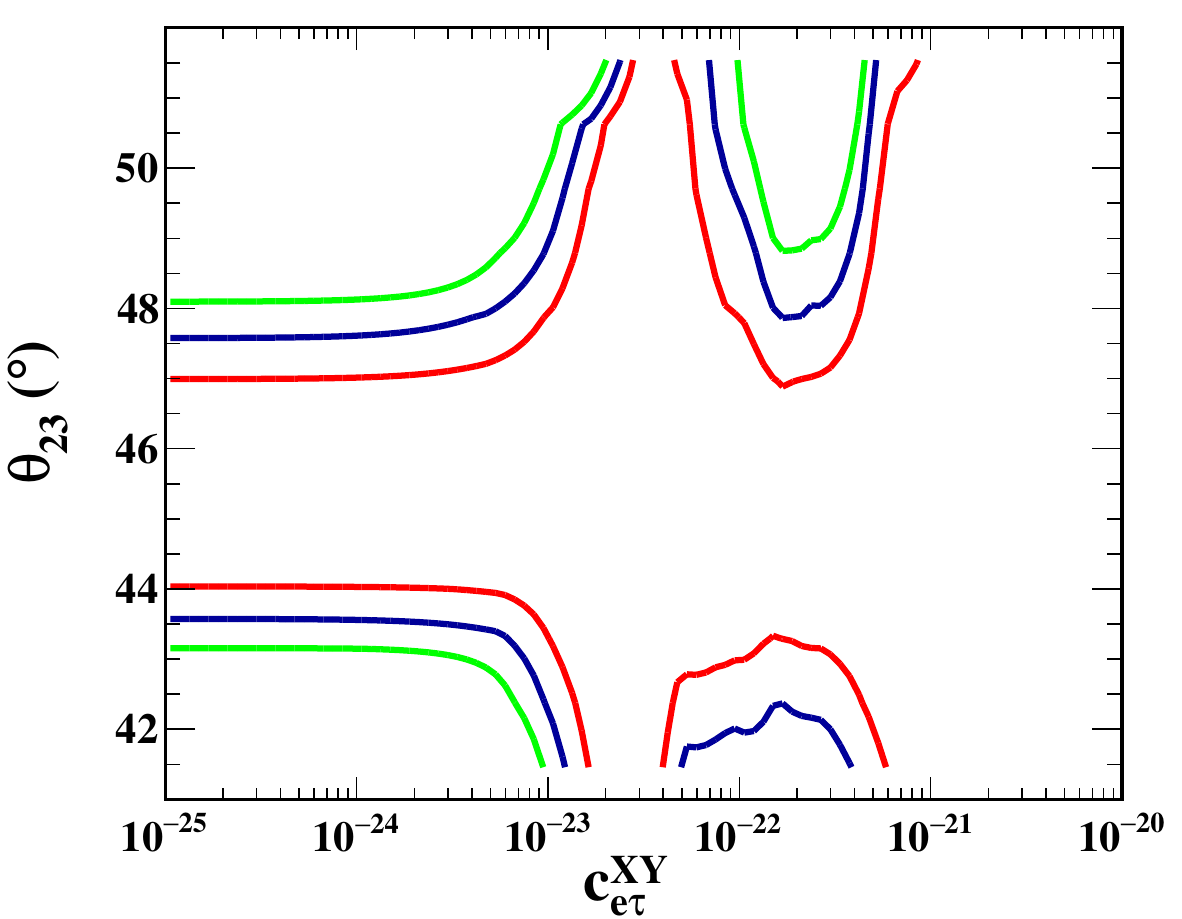}
      \caption{Deterioration of the discovery potential for the $\theta_{23}$ octant as a function of the strength of the true LIV parameter  $\cxyet$.}
     \label{livs}
    \end{figure}
    
      \begin{figure}[!]
      \centering
       \includegraphics[height= 0.25\textwidth,width=0.40\textwidth]{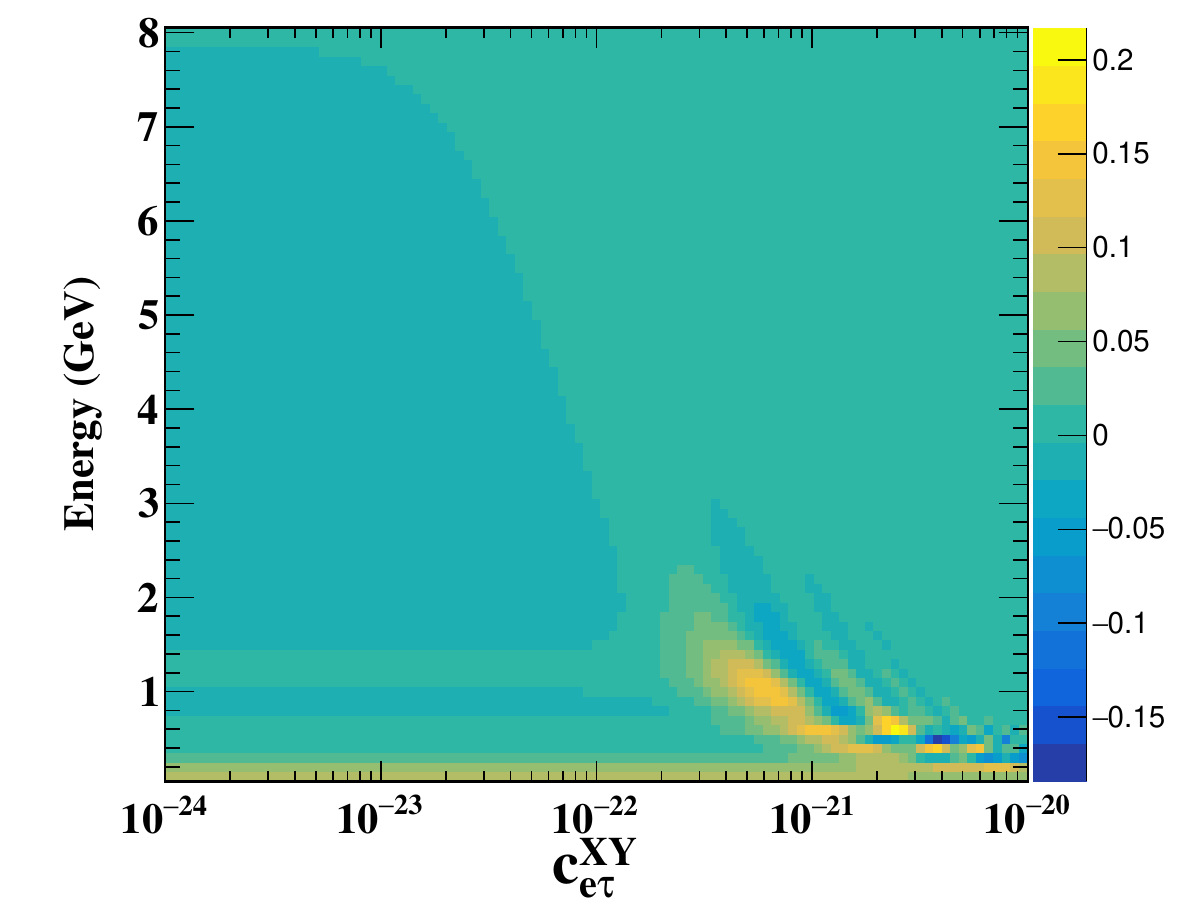}
      \caption{Probability difference ($\Delta P_{\mu e}$) for the opposite value of $\theta_{23}$ in the Energy $-$ $\cxyet$ plane.}
     \label{defprobs}
    \end{figure}
    
We previously reported benchmark results for the Lorentz-violating (LIV) parameter set at $5.0 \times 10^{-23}$. In our current study, we explored how varying the LIV parameter across its allowed range affects the discovery potential of the $\theta_{23}$ octant. For simplicity, we focused on a single parameter, $\cxyet$, which has a significant impact.
Figure~\ref{livs} illustrates the sensitivity to the true $\theta_{23}$ octant as a function of LIV parameter strength, assuming normal ordering (NO) in both data and theory. Our analysis includes marginalization over opposite octant values of the test $\theta_{23}$ including ${45^\circ}$, all phases, $\Delta m^{2}_{31}$ and $\theta_{13}$ are also marginalized across their ranges in both true and test cases. The results, presented at 1$\sigma$ (solid green), 2$\sigma$ (solid blue), and 3$\sigma$ (solid red) confidence levels, indicate that increasing the LIV parameter strength generally reduces the ability to accurately determine the true $\theta_{23}$ octant.
Figure~\ref{defprobs} displays a 2D histogram of the probability difference between the lower octant (LO) ($\theta_{23} = 41^\circ$) and higher octant (HO) ($\theta_{23} = 52^\circ$) cases, assuming normal hierarchy (NH). This histogram is plotted as a function of the LIV parameter strength and energy, with all phases considered as zero.
It is evident from Figure~\ref{defprobs} that around $10^{-22}$, the probability difference increases, which enhances sensitivity in this region (as shown in Figure~\ref{livs}). However, beyond the limit of $10^{-21}$, sensitivity is completely lost.


{\bf {\em Conclusion and discussion:}} 
In conclusion, our study represents the first exploration of the impact of non-isotropic Lorentz invariance violation on identifying the octant of the mixing angle $\theta_{23}$ in the context of DUNE. We have demonstrate that the presence of energy-dependent $\cxyab$ parameters significantly deteriorated the sensitivity of $\theta_{23}$, posing challenges in distinguishing between its octants.
In contrast to the Standard Model scenario, where there is no degeneracy between $\delta_{cp}$ and $\theta_{23}$ over the entire 10-year run period with equal sharing between neutrino and antineutrino modes ~\cite{DUNE:2020jqi}, the introduction of non-isotropic LIV lifts this degeneracy. This effect is particularly evident due to the energy-dependent parameters $\cxyem$ and $\cxyet$. Furthermore, we show that considering the current maximal limits of these parameters exacerbates this issue.

Overall, our findings underscore that non-diagonal LIV has a profound impact on one of the central objectives of the DUNE experiment: the precise determination of the $\theta_{23}$ octant. This presents a significant challenge in achieving accurate measurements of neutrino oscillation parameters and emphasizes the necessity of integrating LIV effects into neutrino oscillation analyses. 
To rigorously investigate the sidereal effect, it is essential to account for variations in the protons on target (POT) throughout the experiment. While we have assumed a constant average POT over the entire sidereal period, any modulation in the average POT with respect to sidereal time can significantly amplify the sidereal effect, potentially leading to substantial deviations from standard expectations even for minimal levels of LIV.
To ensure the success of future precision experiments like DUNE, it is imperative to comprehensively address these LIV effects. This study highlights a critical area for further research and intervention to maintain the integrity of precision neutrino measurements in the presence of LIV.

{\bf {\em Acknowledgments:}} We would like to thank Joachim Kopp for his valuable discussions. S. M., S. S., and V. S. acknowledge financial support from the Department of Science and Technology (DST), New Delhi, India, under the Umbrella Scheme for Research and Development. V.S and L.S. also acknowledge support from  DST-FIST, India. S. S. acknowledges financial support from the Council of Scientific and Industrial Research (CSIR), New Delhi, India. L. S. acknowledges support from the University Grants Commission--Basic Scientific Research Faculty Fellowship Scheme (UGC-BSR) Research Start Up Grant, India (Contract No. F.30-584/2021 (BSR)).
    
\bibliography{Octant_dune}
\end{document}